\documentclass[superscriptaddress,secnumarabic,onecolumn,
amssymb,amsmath,nobibnotes,aps,prd,showkeys,showpacs,nofootinbib]{revtex4}%
\usepackage{graphicx}
\usepackage{epsf}
\usepackage{bm}
\usepackage{amsmath}
\usepackage{amsfonts}
\usepackage{amssymb}
\usepackage{epstopdf}
\usepackage{natbib}%
\usepackage{color}
\providecommand{\U}[1]{\protect\rule{.1in}{.1in}}
\newcommand{\be}{\begin{equation}}
\newcommand{\ee}{\end{equation}}

\newcommand{\mincir}{\raise
-3.truept\hbox{\rlap{\hbox{$\sim$}}\raise4.truept\hbox{$<$}\ }}
\newcommand{\magcir}{\raise
-3.truept\hbox{\rlap{\hbox{$\sim$}}\raise4.truept\hbox{$>$}\ }}

\begin{document}
\title{Bulk viscous quintessential inflation}
\author{Jaume  Haro}
\email{jaime.haro@upc.edu}
\affiliation{Departament de Matem\`atica Aplicada I, Universitat
Polit\`ecnica de Catalunya, Diagonal 647, 08028 Barcelona, Spain}
\author{Supriya Pan}
\email{span@research.jdvu.ac.in}
\affiliation{Department of Physical Sciences, Indian Institute of Science Education and Research Kolkata, Mohanpur $-$ 741246, West Bengal, India}
\affiliation{Department of Mathematics, Jadavpur University, Kolkata $-$ 700032, West Bengal, India}

\begin{abstract}
In a  spatially flat Friedmann-Lema\^itre-Robertson-Walker universe, the incorporation of bulk viscous process in General Relativity leads to
an appearance of a nonsingular background of the universe that both at early and late 
times depicts an accelerated universe.
These early and late scenarios of the universe can be  analytically calculated and mimicked, in the context of General Relativity, by a single scalar field whose potential could also be obtained analytically where the early inflationary phase is described by a one dimensional Higgs potential and the current acceleration is realized by an exponential potential. We show that 
the early inflationary universe leads to a power spectrum of the cosmological perturbations which
match with current observational data, and after leaving the inflationary phase,
the universe suffers a phase transition needed to explain the reheating of the universe via gravitational particle production. 
Furthermore, we find that at late times, the universe enters into the de Sitter phase that can explain the current cosmic acceleration. Finally, we also find that such bulk viscous dominated universe attains the thermodynamical equilibrium, but in an asymptotic manner.  
\end{abstract}
\maketitle
\section{Introduction}

Observational evidences \cite{Riess:1998cb,Perlmutter:1998np, Ade:2015xua} suggest that, currently our universe is experiencing a phase of accelerated expansion. This acceleration is not unique since 
according to both theoretical and observational arguments, our universe had experienced another accelerating phase that occurred during its very early evolution, named as inflation \cite{Guth:1980zm,Linde:1981mu} (see also \cite{Olive:1989nu} and the references therein). The observational results further inform that in between these two accelerating phases separated by a long interval of cosmic time, the intermediate stage of the universe was decelerating and such decelerating phase of the universe had been dominated with different kind of fluids $-$ primarily by relativistic radiation and then by non-relativistic matter. In order to account of such complete universe evolution, attempts have been made in the past either with some matter modifications in general relativity or with different gravity theories other than general relativity. Nevertheless, searching for a unified cosmic evolution still remains as one of the interesting topics in current cosmological research. The motivation of the current work is something similar. Here we consider the bulk viscous process in general relativity and try to see how the universe evolution could be mimicked and moreover how much the scenarios might be viable in respect to the current observational data. 

Recently, this bulk viscous phenomena (see, for instance, \cite{Zimdahl1.1, Zimdahl1.2, Zimdahl1.3, Zimdahl1.4} for a detailed description) got considerable interest since this dissipative mechanism in the Friedmann-Lema{\^\i}tre-Robertson-Walker (FLRW) space-time is able to explain the inflation \cite{bo,bt} and the current cosmic acceleration \cite{beno,begt}.  Moreover, it was found that the cosmology of bulk viscous mechanism can unfiy both accelerating phases respectively in early and late times (see \cite{bcno} for a review),
where essentially the universe starts with an unstable de Sitter solution (early accelerated phase) and ends in a stable one (current accelerating phase). We note that early time acceleration could not always represent an inflationary scenario, for instance, the early time acceleration proposed in current models does not correspond to an inflationary phase \cite{pan1}. 

In this work we have considered a spatially flat FLRW universe 
endowed with bulk viscous mechanism. Additionally, we consider 
a perfect fluid with linear equation of state. 
In connection with that it is 
very reasonable to note that a non-vanishing matter creation rate 
is dynamically equivalent
to an effective bulk viscous pressure but however, both
the formalisms are thermodynamically different \cite{lima1992}. Thus, 
in this work we offer a section on the thermodynamics of the bulk viscous
process.  Our analysis shows that the model provides with 
a nonsingular background (without the big bang singularity) that at early times,
depict an inflationary phase which  ends in a sudden phase transition in order to produce enough particles to reheat the universe,
and at late times, the model renders an accelerated universe. With such nonsingular background, using the reconstruction method, we find the inflationary quintessential potential which 
at early times, predict a spectral index of scalar cosmological perturbations with running and its corresponding ratio of tensor to scalar perturbations that fit well with recent observational data.
Moreover, these potentials have an absolute minimum at the corresponding de Sitter solution, meaning that the de Sitter solution is a late time  attractor, and thus,  the
background will depict the current cosmic acceleration. 

We have organized the work in the following way: In section \ref{sec-BV} we have shortly described
the bulk viscous cosmology in the flat FLRW universe. Then in section \ref{sec-model}, introducing the
model, we analytically solved the background and discussed its dynamical behavior. Further, in section \ref{sec-fieldTH}, we explicitly describe the equivalent field theoretic description of the bulk viscous universe and elaborately discussed the inflationary phase. 
Next in section \ref{thermo}, we have shown the present model is consistent with the thermodynamics of the universe. Finally, we have summarized our results in section \ref{summary}. 

We note that the units used throughout the paper have been set as $\hbar=c=8\pi G=1$.

\section{Bulk viscous cosmology}
\label{sec-BV}
In this section we shall describe the gravitational equations for a bulk viscous universe in a spatially flat  Friedmann-Lema\^itre-Robertson-Walker (FLRW) geometry. The line element for the FLRW geometry is given by
$$ds^2= -dt^2+ a^2(t) \Bigl[ dr^2+ r^2 \left(d \theta^2 + \sin^2 \theta \, d \phi^2\right)  \Bigr],$$
where $a(t)$ is the expansion scale factor of the universe. 
In cosmology, the simplest effective way to incorporate the
bulk viscosity is to use Eckart theory \cite{eckart} (see also \cite{Zimdahl1.1, Zimdahl1.2, Zimdahl1.3, Zimdahl1.4}),
in which the pressure $p$ is actually replaced by $\bar{p}= p- 3 H \xi(t)$, where $\xi(t) > 0$, is the  so-called
coefficient of bulk viscosity \cite{bg}, which in general could be a function 
of any cosmic variable or constant. 
Now, in this flat FLRW spacetime,
the  classical Friedmann and Raychaudhuri's equations are modified as
\begin{eqnarray}
\rho = 3H^2,\quad
\dot{H}=- \frac{1}{2} (p+\rho)+ \frac{3}{2} H \xi(t),\label{bulk-friedmann2}
\end{eqnarray}
where $\rho$ is the energy density of the universe. One can show that the inclusion of such bulk viscosity is equivalent to an inhomogeneous equation of state that 
can be derived from the context of general relativity, or modified gravity, 
for instance the $f(R)$ gravity or even from the braneworld scenario, and consequently, one finds that the inhomogeneous 
equation of state takes the form \cite{Nojiri:2005sr, Capozziello:2005pa, Nojiri:2006zh, bcno}

\begin{eqnarray}
\bar{p} = p - g (t, a(t), H, \dot{H},...)
\end{eqnarray}
where $g (t, a(t), H, \dot{H},...)$ is any arbitrary continuous function of the prescribed variables. One can see that for $g (t, a(t), H, \dot{H},...) = 3 H \xi (t)$, the bulk viscous cosmology is 
recovered. However, there is no such specific rule for the choices for the bulk visocus coefficient $\xi (t)$, or equivalently, the function $g(t, a(t), H, \dot{H},...)$. We recall some earlier as
well as latest investigations \cite{Barrow:1986yf, Barrow:1988yc, bo, bt, Brevik:2016kwq} where the inflationary solutions have been discussed for different choices for $\xi (t)$. In addition, the 
late cosmic acceleration can also be achieved in this context \cite{beno, begt, bcno}. For a comprehensive discussions on the cosmology of bulk viscous models, we refer to a recent 
review \cite{Brevik:2017msy}. In principle, although the models for the bulk viscous cosmology are motivated from the phenomenological ground, but its equivalence to an inhomogeneous equation of state makes it appealing. In this work we shall 
restrict ourselves to $\xi (t) \equiv \xi (H)$, that means we shall neglect the possibilities of the derivative terms of $H$. \\
 
Now, assuming that the universe is filled with a  barotropic fluid with Equation of State (EoS): $p= (\gamma- 1)\rho$ ($1 \leq \gamma \leq 2$),
where the dimensionless parameter $\gamma$ has been chosen to be greater or equal to $1$, in order to have a non-negative pressure. Hence, Raychaudhuri's equation can be written as
\begin{eqnarray}
\dot{H} = -\frac{3}{2}(1+w_{eff}(H)) H^2\label{bulk-friedmann3},
\end{eqnarray}
where we have introduced  the effective EoS parameter as
\begin{eqnarray}
w_{eff}(H)&\equiv& -1-\frac{2\dot{H}}{3H^2}= -1+ \gamma \left(1-\frac{\xi(H)}{\gamma H}\right).\label{eff-eos}
\end{eqnarray}

Note that, equation (\ref{bulk-friedmann3}) is equivalent to the standard Raychaudhuri's equation for a non-linear EoS of the form
\begin{eqnarray}
 p=(\gamma-1)\rho-\sqrt{3\rho}\xi\left(\sqrt{\frac{\rho}{3}}\right).
\end{eqnarray}

However, one may find the nature of $w_{eff}(H)$ after taking a cosmic time differentiation as
\begin{align}
\dot{w}_{eff} (H) & = -\frac{3\gamma}{2} \left(1-\frac{\xi(H)}{\gamma H}\right) \Bigl[\xi (H)- H \xi^{\prime} (H)\Bigr]
\end{align}
where the prime stands for the differentiation with respect to $H$. So, $w_{eff}$ could be an increasing function or a decreasing
function depending on the nature of the terms. We note that for $\xi (H)= $constant, say $\xi_c $, the nature of the function $w_{eff}$ is completely
determined. 

Now in order to account for both accelerating phases, 
it is reasonable to ask whether $\xi (H)$
may allow two de Sitter solutions where necessarily 
one of them will be an attractor and the other one must be a repeller. 
Precisely, whatever the form for $\xi (H)$ is chosen, the equation $1+w_{eff}(H)=0$ should have two positive roots, namely
$H_{+}>H_{-}$, that will correspond to
different de Sitter solutions. Effectively, when $1+w_{eff}(H)=0$, the equation  (\ref{bulk-friedmann3}) becomes $\dot{H}=0$, meaning that the roots $H_{\pm}$ of $1+w_{eff}(H)=0$, correspond to  the solutions  $H(t)=H_{\pm}$ of (\ref{bulk-friedmann3}), and thus, for these solutions the scale factor becomes $a_{\pm}(t)=a_{\pm} e^{H_{\pm}t}$, which depict two
de Sitter solutions. When $H_+$, which eventually could be $+\infty$, is the
repeller and $H_-$ is an attractor,
we will have a nonsingular background
with $w_{eff}(H)> -1$, that starts at $H_+$ and ends at $H_-$, depicting an accelerated universe at early and late times.  This is a candidate to depict our universe.
However, to check its
viability, one has to show that at early times, this accelerated phase is an inflationary one. That means we need to check the viabilities of the model from the 
observational data. \newline

\section{The model and its dynamical analysis}
\label{sec-model}

As dicussed in section \ref{sec-BV}, the equivalence between bulk viscosity and the inhomogeneous equation of state enables us
to consider several phenomenological expressions to construct different universe models. On the other hand, the bulk viscous process is equivalent to an open 
cosmology in presence of gravitational particle production \cite{Prigogine98}. We make a note that, although the gravitational
equations for both bulk visocus models and particle creation models are equivalent but both the processes are thermodynamically different \cite{lima1992}, as already mentioned. We consider a
typical choice for the bulk viscous coefficient which has the following expression \cite{pan}
\begin{eqnarray}\label{bv}
 \xi(H)=-\xi_0+m\,H+\frac{n}{H},
\end{eqnarray}
where $\xi_0$, $m$ and $n$ are some positive parameters with the condition that $\xi (H)> 0$ which is a consequence of the 
entropy growth in irreversible processes of energy dissipation (see section $49$ of \cite{ll}). It is  clear that the above form of the bulk viscous coefficient (\ref{bv}) represents a
generalized model since the particular models, namely, $\xi (H)= $ constant \cite{beno}, $\xi (H) \propto H$ \cite{beno} and for $\xi (H) \propto 1/H$, can be recovered as special cases.  Now, 
using the condition of $\xi (H) > 0$, then taking into account its minimum achieved at $\sqrt{\frac{n}{m}}$, its positivity condition is satisfied when $\frac{\xi_0^2}{4m}\leq n$. We now concentrate on the dynamical analysis of the model (\ref{bv}). Using the Friedmann equations in (\ref{bulk-friedmann2}) for this bulk viscous model, one arrives at

\begin{eqnarray}\label{sp-bv}
\dot{H} = F (H) = \frac{3}{2} \Bigl[ (m -\gamma) H^2 - \xi_0 H + n \Bigr]
\end{eqnarray}
which is a  first order differential equation and the dynamics of the universe can be obtained from the analysis of the fixed points of this equation. Interestingly enough, the 1-dimensional system characterized by Eq. (\ref{sp-bv}) is a property that appears also in $f(T)$ gravity, see \cite{Awad:2017yod} (especially Eq. (3.1) of \cite{Awad:2017yod}).
Now, solving the equation (\ref{sp-bv}),  for $\dot{H} =0$, one can 
obtain the fixed points of (\ref{sp-bv}).  That means, if $H_{\ast}$ is a fixed point of (\ref{sp-bv}), then $F (H_{\ast}) = 0$, i.e. $H_{\ast}$ is a root of $F (H)$. At the nonzero fixed points, it 
is evident that the FLRW metric describe the de Sitter universes. Now, the stability of a particular fixed  point $H_{\ast}$ depends on the sign of the quantity $dF (H)/dH$ calculated at 
that particular point. More precisely, if $dF (H)/dH < 0$ at $H = H_{\ast}$, then $H_{\ast}$ is asymptotically stable, also known as attractor, meaning that all the solutions near $H = H_{\ast}$
approach asymptotically to it at very late times, while on the
other hand if $dF (H)/dH > 0$ at $H = H_{\ast}$, then the fixed point is unstable and it is known as a repeller, meaning that at very early times all the solutions near $H = H_{\ast}$ move away from it. Now 
for the current model in (\ref{bv}), 
solving the  dynamical equation (\ref{sp-bv}), the fixed points, i.e.,  the de Sitter solutions are 

\begin{eqnarray}\label{critical-points}
 H_{\pm}=\frac{\xi_0}{2(m-\gamma)}\left(1\pm \sqrt{1+\frac{4(\gamma-m)n}{\xi_0^2} }\right).
\end{eqnarray}
One can notice that in order to have two real fixed points, we must have 
$1+\frac{4(\gamma-m)n}{\xi_0^2} > 0$ and indeed $H_{\pm} > 0$, since we have considered 
the expanding branch of the universe. To perform the dynamical analysis we consider the first possibility of $\xi_0 > 0$ and we have the following regions namely,
\begin{enumerate}
 \item  [(i)] $\mathcal{R}_1 = \{(m, n): m-\gamma \geq 0, n> 0, 1+\frac{4(\gamma-m)n}{\xi_0^2} > 0 \} $ where $0 < H_{-} < H_{+}$. Here $H_{+}$ is a repeller while 
$H_{-}$ is an attractor.

\item  [(ii)] $\mathcal{R}_2 = \{(m, n): m-\gamma \leq 0, n < 0, 1+\frac{4(\gamma-m)n}{\xi_0^2} > 0 \} $ where $ H_{+} < H_{-} < 0$. Here $H_{+}$ is a 
repeller while $H_{-}$ is an attractor.
\end{enumerate}

\textbf We also consider the second possibility when $\xi_0 <0$. In that case we have the following regions:

\begin{enumerate}
 \item [(iii)] $\mathcal{R}_3 = \{(m, n): m-\gamma \geq 0, n> 0, 1+\frac{4(\gamma-m)n}{\xi_0^2} > 0 \} $ where $ H_{+} < H_{-} < 0$. Here, $H_{+}$ is an
attractor and $H_{-}$ is a repeller.

\item [(iv)] $\mathcal{R}_4 = \{(m, n): m-\gamma \leq 0, n < 0, 1+\frac{4(\gamma-m)n}{\xi_0^2} > 0 \} $ where $0 < H_{-} < H_{+}$. Here $H_{+}$ is 
an attractor while $H_{-}$ is a repeller. 
\end{enumerate}
 
From this analysis, we can infer that the fixed points for which one gets a nonsingular dynamics without the big bang singularity in addition to have
two accelerated phases of the universe, one at early times, and one at late-times (current accelerating phase), we must have two fixed points $H_{+}$ and $H_{-}$ with
$0 < H_{-} < H_{+}$ so that $H_{+}$ is a repeller and $H_{-}$ becomes an attractor. Thus, one can see that such possibility may happen if 
parameters $\xi_0$, $m$ and $n$  belong to the following domain of $\mathbb{R}^3$ 
\begin{eqnarray}\label{eqn-W}
W=\{\xi_0>0, m\geq \gamma, n>0, \mbox{and}~\frac{4(\gamma-m)n}{\xi_0^2}>-1 \}.
\end{eqnarray}

Moreover, since at the critical points $w_{eff}=-1$, one will have
a universe starting and ending in an accelerated phase. This is the main result of this work since the mechanism of bulk viscous cosmology in the flat FLRW universe can 
successfully generate two successive accelerating phases in a huge time gap with a nonsingular background.

Assuming that, $n\ll \xi_0^2\min\left(m-\gamma,\frac{1}{m-\gamma} \right)$, one obtains
$H_+\cong \frac{\xi_0}{m-\gamma}$, $H_-\cong \frac{n}{\xi_0}$, with $0<H_-\ll H_+$, and consequently, the universe could start at very high energies and ends at low ones. In particular,
for $m=\gamma$, one has $H_+=\infty$,
but there is no big bang singularity (a singularity at early finite time), because in that case, $\dot{H}=-\frac{3}{2}(\xi_0H-n)$, {{}which for large values of $H$ gives}
\begin{eqnarray}
 \dot{H}\cong -\frac{3}{2}\xi_0H\Longleftrightarrow H(t)\cong H_ie^{-\frac{3}{2}\xi_0(t-t_i)},
\end{eqnarray}
meaning that $H$ diverges only when $t=-\infty$.

\begin{figure}
\includegraphics[width=0.40\textwidth]{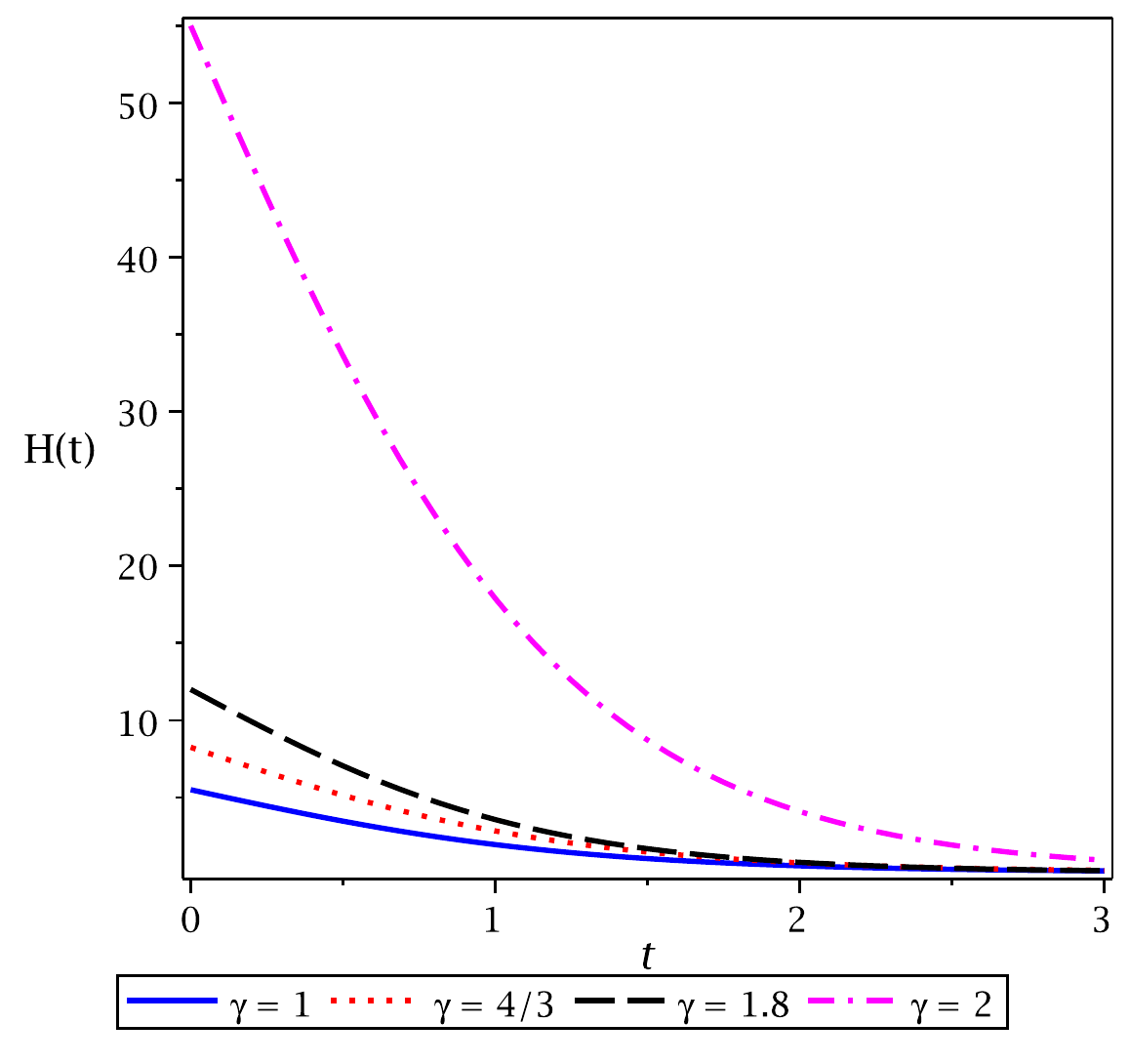}
\includegraphics[width=0.42\textwidth]{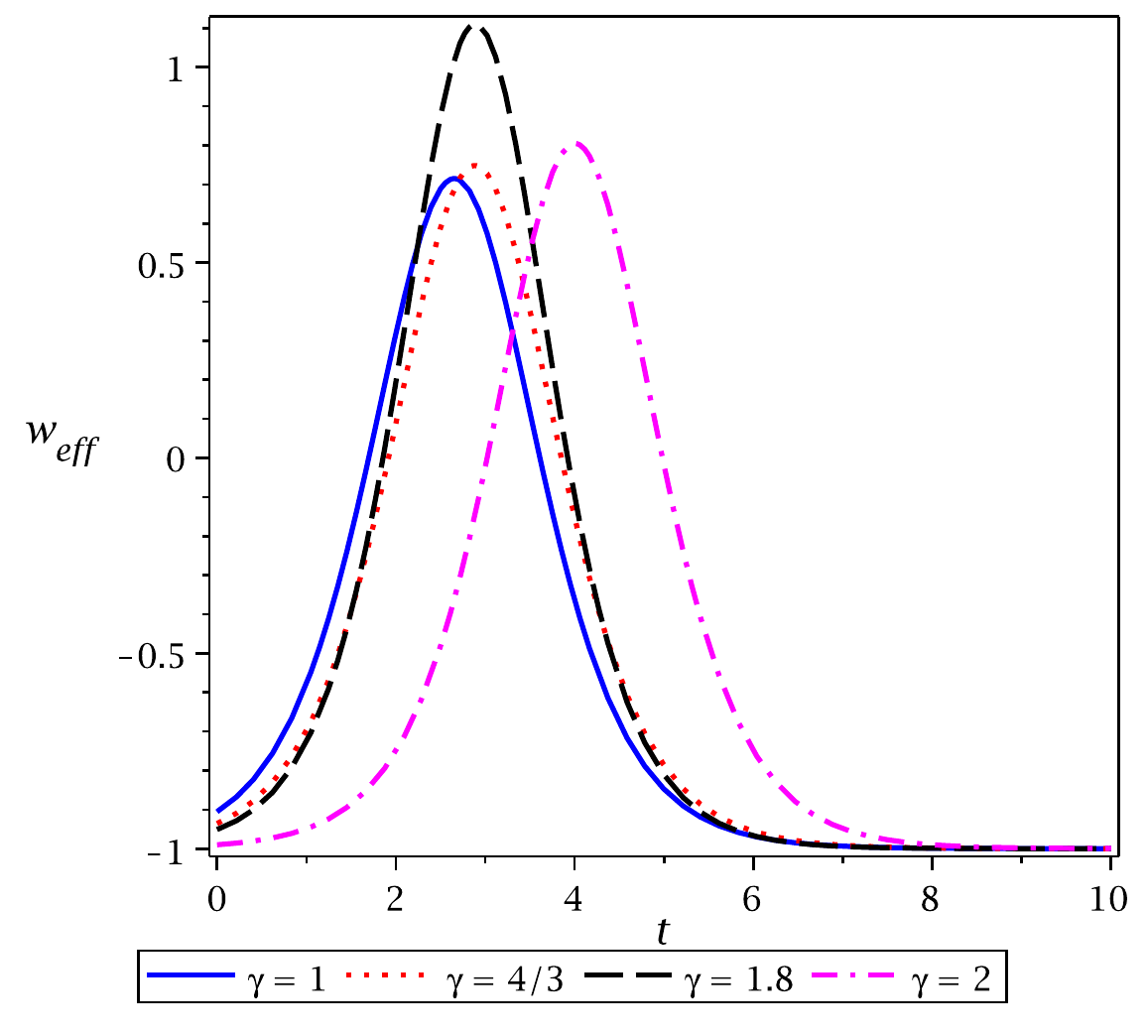}
\caption{The evolutions for the Hubble rate (\ref{hubble}) and the effective equation of state for this Hubble rate have been shown for different values of $\gamma$, respectively 
in the left and right panels of this figure. The common parameters belong to $W$ of (\ref{eqn-W}). One may notice that the effective equation of state 
shown in the right panel of this figure starts from $w_{eff}= -1$ at the very early evolution of the universe and as $t$ increases, it approaches toward the cosmological constant boundary.} 
\label{figure-Hubble-EoS}
\end{figure}

In fact, for our model, the Raychaudhuri equation (\ref{bulk-friedmann3}) can be analytically solved for $m>\gamma$ as

\begin{eqnarray}\label{hubble}
 H(t)=\frac{H_+e^{-\frac{3}{4}(H_+-H_-)(m-\gamma)t}+
 H_-e^{\frac{3}{4}(H_+-H_-)(m-\gamma)t}}{2\cosh{\left(\frac{3}{4}(H_+-H_-)(m-\gamma)t\right)}},
\end{eqnarray}
and
\begin{eqnarray}\label{hubble1}
 H(t)=\xi_0e^{-\frac{3}{2}\xi_0 t}+\frac{n}{\xi_0},
\end{eqnarray}
for $m=\gamma$, and it is evident that for $t\rightarrow \infty$, $H$ tends to a constant value, that means an exponential expansion should happen. In fact, from equation (\ref{hubble}),
since $\lim_{t\rightarrow \pm\infty}=H_{\mp}$, we can deduce that the Hubble parameter evolves from $H_+$ to $H_-$, i.e. at very early times the universe leaves
the de Sitter phase with $H=H_+$ and at late times enters, asymptotically,  once again in a de Sitter phase given by $H=H_-$,
and the same happens with the equation (\ref{hubble1}), where the Hubble parameter
evolves from $\infty$ to $\frac{n}{\xi_0}$. In the left panel of Fig. (\ref{figure-Hubble-EoS}), we show the evolution of the Hubble parameter (\ref{hubble}) for different values 
of $\gamma$ with the parameters from $W$ in (\ref{eqn-W}). The right panel of Fig. \ref{figure-Hubble-EoS} shows  the effective equation of state for this Hubble rate using the 
same values of $\gamma$ and the parameters from the same domain. One can see that the effective equation of state (right panel of Fig. \ref{figure-Hubble-EoS}) starts at very late times ($t=-\infty$)
from $w_{eff}= -1$ and it approaches  asymptotically at late times toward $w_{eff} = -1$. 

In the following we shall establish an equivalent field theoretic description of this 
bulk viscous universe aiming to account of both early and late accelerated universe 
through a single scalar field potential.


\section{Inflationary quintessential potential: One potential connecting two distinct phases}
\label{sec-fieldTH}

In this section we shall consider a scalar field $\varphi$ with 
potential $V (\varphi)$ that is minimally coupled to gravity and 
then and we will see under which conditions a scalar field could 
mimick the dynamics of a perfect fluid with bulk viscosity 
in order to provide viable backgrounds that could depict our 
universe correctly \footnote{We note that hydrodynamical
perturbations is not compatible to 
provide a suitable description for primeval perturbations ((see \cite{mfb} for a detailed discussion)) since during the inflationary period, one has $\bar{p}=(\gamma-1)\rho-3H\xi(H)\cong-\rho$, and thus, the square of the velocity of sound, which appears in the Mukhanov-Sasaki equation
\cite{ms1, ms2}, could approximately becomes $c_{s}^{2}\equiv\frac{\dot{p}}{\dot{\rho}%
}\cong-1$. Thus, Jeans instability appears for modes well
inside the Hubble radius. But for scalar field theory, one always has $c_{s}^{2}=1$.
This is an essential reason why the dynamics of a hydrodynamical fluid is tried 
to mimick with scalar field descriptions. }. 
Thus, in a flat FLRW universe, the energy density and the pressure of the scalar field
are repsectively calculated as  $\rho_{\varphi}= \frac{1}{2} \dot{\varphi}^2+ V (\varphi),$
$p_{\varphi}= \frac{1}{2} \dot{\varphi}^2- V (\varphi)$.  To show the  equivalence with the bulk viscous system (\ref{bulk-friedmann2}), we perform the replacement $\rho\longrightarrow \rho_{\varphi}$, and $ p-3H\xi(H) \longrightarrow p_{\varphi}$,
to recover the standard Friedmann and Raychaudhuri equations for a universe filled by an scalar field
\begin{eqnarray}
 3H^2=\rho_{\varphi}, \quad 2\dot{H}=-\dot{\varphi}^2.\label{friedmann2}
\end{eqnarray}

Note that, since the equations in (\ref{friedmann2}) are the usual equations for a single scalar field, that means the dynamics driven by a fluid with bulk viscosity with
an effective EoS parameter greater than $-1$, could be mimicked by
a single scalar field in the context of General Relativity.

Combining  the Friedmann and Raychaudhuri equations  (\ref{bulk-friedmann2}), (\ref{bulk-friedmann3}) and  (\ref{friedmann2}), one easily obtains
\begin{eqnarray}
\dot{\varphi}= \sqrt{-2\dot{H}}=\sqrt{3\gamma H^2 \left(1- \frac{\xi(H)}{\gamma H}\right)}~,\label{scalarfield2}\\
V (\varphi)= \frac{3H^2}{2}\left[(2- \gamma)+ \frac{\xi(H)}{ H}\right] ~.\label{potential}
\end{eqnarray}

Now, performing the change of variable $dt=\frac{dH}{\dot{H}}$, the equation (\ref{scalarfield2}) becomes $\varphi=-\int\sqrt{-~\frac{2}{\dot{H}}}~dH =-\frac{2}{\sqrt{3}}\int \frac{dH}{\sqrt{\gamma H^2-\xi(H)H}}$, which for our model (\ref{bv}) turns out to be 
\begin{eqnarray}
 \varphi=-\frac{2}{\sqrt{3}}\int \frac{dH}{\sqrt{(\gamma-m)H^2+\xi_0H-n}},
\end{eqnarray}
and when the parameters belong to the domain $W$, for $m<\gamma$ one has (see formula $2.261$ of \cite{gr})
\begin{eqnarray}\label{sinus}
  \varphi=\frac{2}{\sqrt{3(m-\gamma)}}\arcsin\left(\frac{2(\gamma-m)H+\xi_0}{\sqrt{\xi_0^2+4(\gamma-m)n}}\right),
\end{eqnarray}
defined in $\left( - \frac{\pi}{\sqrt{3(m-\gamma)}}, \frac{\pi}{\sqrt{3(m-\gamma)}}\right)$, and when $m=\gamma$
\begin{eqnarray}
 \varphi=-\frac{4}{\sqrt{3}\xi_0}\sqrt{\xi_0H-n},
\end{eqnarray}
which is defined in the domain $(-\infty,0)$.

Finally, isolating $H$ and inserting it in (\ref{potential}), one obtains the corresponding potentials. Once the  potential has been reconstructed,  one has the corresponding
conservation equation
\begin{eqnarray}\label{KG}
 \ddot{\varphi}+\sqrt{3}\sqrt{\frac{\dot{\varphi}^2}{2}+V(\varphi)}\dot{\varphi}+V_{\varphi}(\varphi)=0,
\end{eqnarray}
which is a second order differential equation,  whose infinitely many solutions depict different backgrounds, and where one of them is the  solution of (\ref{bulk-friedmann3}).

For example,
in the simplest case $m=\gamma$, one easily obtains
\begin{eqnarray}\label{example}
 V(\varphi)=\frac{27}{256}\xi_0^2\varphi^4+\frac{9}{8}\left(n-\frac{\xi_0^2}{4} \right)\varphi^2+\frac{3n^2}{\xi_0^2}.
\end{eqnarray}

Recall that, the positivity of $\xi(H)$ implies $\frac{\xi_0^2}{4\gamma}\leq n$, and
note that,  when $n\geq\frac{\xi_0^2}{4}$, the unique minimum of the potential (\ref{example}) is {{}achieved} at $\varphi=0$. On the contrary, when $n<\frac{\xi_0^2}{4}$, the potential has a maximum at
$\varphi=0$. We also add that for $H_-$ to be an attractor, we should have $n\geq\frac{\xi_0^2}{4}$. Moreover, as the fluid has positive pressure
($m=\gamma\geq 1$), this condition is compatible with the
positivity of $\xi(H)$ because $\frac{\xi_0^2}{4\gamma}\leq \frac{\xi_0^2}{4}$, {{}which} means that the critical point satisfies $H_-=\frac{n}{\xi_0}\geq \frac{\xi_0}{4}$. Unfortunately,
the model we have chosen
has two important defaults concerning the slow roll parameter \footnote{\textit{As we well see,  the main slow-roll parameter is of the order $\frac{\xi_0}{H}$, this means that,
 in order to match with current observational data,
the observable modes must leave the Hubble radius at scales of the order $H\sim 10^{3}\xi_0\leq 10^3 H_-$. However, since $H_-$ has to be close to the current value of the Hubble parameter, the condition $H\leq 10^3 H_-$ is compatible with the fact that the observational modes must
leave the Hubble radius at
high energy densities (few orders below Planck's one).}}, and about the reheating of the universe \footnote{\textit{There is no mechanism to reheat the universe, because neither oscillations nor abrupt phase transitions at high scales, that breakdown the adiabaticity to produce enough amount of
particles that thermalize the universe after inflation, occur in that models.}}.

However, the problmes can be easily overpassed with the introduction of a phase transition at early times when the universe ceases to accelerate. More precisely, we  will choose
the phase transition when the universe is
stiff matter dominated, i.e., it enters in a deflationary phase \cite{spokoiny}

Taking $\gamma=2\Longleftrightarrow p=\rho$, $n=\frac{\xi_0^2}{8}\Longrightarrow \xi(H)\geq 0$, our  continuous coefficient of bulk viscosity is improved
as follows
\begin{eqnarray}\label{model1}
 \xi(H)=\left\{\begin{array}{ccc}
                -\xi_0+2H+\frac{\xi_0^2}{8H},& \mbox{for}& H\geq H_E\\
                \xi_1,&\mbox{ for}& H\leq H_E,
               \end{array}\right.
\end{eqnarray}
where $0<\xi_1\ll \xi_0$, and $H_E=\frac{1}{4}(\xi_1+\xi_0)\left(1+\sqrt{1-\frac{\xi_0^2}{(\xi_1+\xi_0)^2}}  \right)\cong \frac{\xi_0}{4}.$ \ The corresponding potential has the form
\begin{eqnarray}\label{potential3}
 V(\varphi)=\left\{\begin{array}{ccc}
  \frac{27 \xi_0^2}{256} \left( \varphi^2- \frac{2}{3}  \right)^2,&\mbox{for}& \varphi\leq \varphi_E  \\
  \frac{3}{2}H(\varphi)\xi_1,&\mbox{for}& \varphi\geq \varphi_E,
                   \end{array}\right.
\end{eqnarray}
where $\varphi_E=-\sqrt{\frac{2}{3}}\sqrt{\frac{8H_E}{\xi_0}-1}\cong -\sqrt{\frac{2}{3}}~$, and
\begin{eqnarray}
 H(\varphi)=\frac{A^2e^{-\sqrt{\frac{3}{2}}(\varphi-\varphi_E)}+  \xi_1^2e^{\sqrt{\frac{3}{2}}(\varphi-\varphi_E)}}{8A}+\frac{\xi_1}{4}\nonumber\\
 \cong \frac{\xi_0}{4}e^{-\sqrt{\frac{3}{2}}\varphi+1}+ \frac{\xi_1^2}{16\xi_0}e^{\sqrt{\frac{3}{2}}\varphi+1}+\frac{\xi_1}{4},
\end{eqnarray}
with $A\equiv 2\sqrt{4H_E^2-2\xi_1H_E}+4H_E-\xi_1\cong 2\xi_0$.
\textit{It should be noted that eq. (\ref{potential3}) depicts an one dimensional Higgs potential (also known as Double Well Inflationary potential \cite{DWI-A, DWI-B}) which describes an early inflationary era plus an exponential potential responsible for the current accelerating phase.}

Now, the corresponding conservation equation (\ref{KG}) provides  backgrounds that could depict our universe, and one of them is the solution of the Raychaudhuri equation (\ref{bulk-friedmann3})
\begin{eqnarray}\label{nonsingular}
 H(t)=\left\{\begin{array}{ccc}
    \left(H_E-\frac{\xi_0}{8}\right)e^{-\frac{3\xi_0}{2}(t-t_E)} +   \frac{\xi_0}{8}~, &\mbox{for}& t\leq t_E, \\
     \frac{\frac{\xi_1}{2}H_E}{H_E-(H_E-\frac{\xi_1}{2} )e^{-\frac{3\xi_1}{2}(t-t_E)}    }~, &\mbox{for}& t\geq t_E,
      \end{array}\right.
\end{eqnarray}
where $t_E$ is the phase transition time. One can solve the scale factor as 

\begin{eqnarray}\label{nonsingular-scale-factor}
 a(t)=\left\{\begin{array}{ccc}
   a_E e^{\frac{\xi_0}{8} (t-t_E)} e^{-\frac{2}{3\xi_0} \left(H_E -\frac{\xi_0}{8}\right)\left[e^{-\frac{3\xi_0}{2} (t-t_E)}-1\right]}~, &\mbox{for}& t\leq t_E, \\
     a_E \Bigl| \frac{H_E e^{\frac{3\xi_1}{2} (t-t_E)}-\left(H_E -\frac{\xi_1}{2} \right)}{\xi_1/2} \Bigr|~, &\mbox{for}& t\geq t_E,
      \end{array}\right.
\end{eqnarray}

Finally,  at late time, one can see that the system has a critical point at $H_-=\frac{\xi_1}{2}\Longleftrightarrow \varphi=\varphi_E+\sqrt{\frac{2}{3}}\ln\left(\frac{A}{\xi_1}\right)$. It is easy to see that $H_{-}$ is an attractor because, applying the chain rule to
$V_{\varphi\varphi}$ at this critical point, one finds that
\begin{eqnarray}\label{model}
 V_{\varphi\varphi}=-\frac{3\xi_1}{2}\frac{\varphi_{HH}}{\varphi_H^3}=\frac{9}{16}\left(4H_{-}-\xi_1\right)=\frac{9\xi_1}{16}>0,
\end{eqnarray}
which implies that the potential has a minimum at the critical point, and consequently, it is an attractor.

The following remark is in order: When one considers the case $m\gtrsim \gamma=\frac{4}{3}$, and assumes a  phase transition as in the model (\ref{model1}),
the bulk viscous Raychaudhuri equation ({\ref{bulk-friedmann3}) leads to a nonsingular solution that starts at  $H_+\cong \frac{\xi_0}{m-\frac{4}{3}}$
and ends at $H_-=\frac{\xi_1}{2}$. Then, the corresponding quintessential inflationary potential will have a maximum at $H_+$ (unstable) and a minimum at $H_-$ (attractor), that is,
some backgrounds (models which, at early time, are close to our nonsingular background (\ref{nonsingular}))
given by  (\ref{KG}),  leave the de Sitter phase $H_+$ at early-times, and suffer a sudden phase transition when the universe
starts to decelerate, and finally, enters {{}into the stable} de Sitter phase $H_-$. In fact, the shape of the potential can easily be imagined: From equation
(\ref{sinus}), one can deduce that, before the
phase transition ($\varphi<\varphi_E$), the potential has
a sinusoidal form with period $\frac{4\pi}{\sqrt{3m-4}}$, and after the phase transition, it has the same shape as (\ref{potential3}).\newline

\


As during the early evolution of the universe, our background (\ref{nonsingular})
satisfies $w_{eff}(H)\cong -1$ (i.e. quasi de Sitter), thus, in order to check whether 
such background leads to a power spectrum of cosmological perturbations that fit well with current observational data \cite{Ade}, we introduce the slow roll parameters \cite{btw}:
\begin{eqnarray}
\epsilon=-\frac{\dot{H}}{H^2}, \quad \eta=2\epsilon-\frac{\dot{\epsilon}}{2H\epsilon},
\end{eqnarray}
that allow us to calculate the spectral index ($n_s$), its running ($\alpha_s$) and the ratio of tensor to scalar perturbations ($r$), respectively as, 
$n_s-1=-6\epsilon+2\eta,~\alpha_s=\frac{H\dot{n}_s}{H^2 +\dot{H}},~r=16\epsilon$.  At early times, i.e., when $H>H_E$, introducing 
the notation $x\equiv \frac{3\xi_0}{2H}$, one has $\epsilon=x\left(1-\frac{x}{12}\right),\quad \eta=\epsilon+\frac{x}{2}$,
and as a consequence, $
n_s-1=-3x+\frac{x^2}{3}.$
Conversely, $x=\frac{9}{2}\left(1-\sqrt{1-\frac{4(1-n_s)}{27}}\right)$. Then, given the observational values of the spectral index, one can obtain the range of $x$. 
From the estimation of the inflationary parameters by Planck's team
(see table $5$ of \cite{Ade}), we find the spectral index at 1$\sigma$ Confidence Level (C.L.) to be $n_s=0.9583\pm 0.0081$, which means that, at $2\sigma$ C.L., one
has $0.0085\leq x \leq 0.0193$, and thus, $0.1344\leq r=16\epsilon\leq 0.3072$.}
We notice that when the running is disregarded, the tensor-to-scalar ratio, $r$, for this model is large in compared to the latest observations  from Planck 2013 ($r <0.11$ at 95.5\% CL) \cite{Ade} and 
Planck 2015 ($r< 0.12$ at 95.5\% CL) \cite{Ade:2015lrj}. 
However, our model (\ref{nonsingular})  leads to a negative theoretical value of the running $\alpha_s\cong -\frac{3x\epsilon}{1-\epsilon}$, which at the scales we are dealing with, is constrained to be $-7\times 10^{-4}\leq \alpha_s\leq -2\times 10^{-4}$, entering in the 
1-dimensional marginalized $95.5\%$ C.L., because the observational data \cite{Ade} provide $\alpha_s=-0.021\pm 0.012$. Therefore, when the running is allowed, the 
PLANCK+WP 2013 (see table $5$ of \cite{Ade})  data  constrain
$r\leq 0.25$ at $95.5 \%$ C.L., thus, when $0.0085\leq x\leq 0.0156$, our model leads to
a spectral index and  running belonging to the 1-dimensional marginalized $95.5\%$ C.L., and to a theoretical value of the ratio of tensor to scalar perturbations satisfying
$r\leq 0.25$ at $95.5\%$ C.L.
One can see that $\epsilon = 1$, can be treated as the phase when the universe is just about to start its decelerating phase from the inflation \footnote{Using the definition of the slow roll parameter $\epsilon$, the effective  equation (\ref{eff-eos}) becomes $w_{eff}(H)=-1+\frac{2}{3}\epsilon$. Now, if one assumes that for $\epsilon=1$, slow roll ends, that means alternatively we have $w_{eff}(H_{end})=-\frac{1}{3}$,  where $H_{end}$ be the value of the Hubble parameter when the slow roll ends, then it further means that  
the condition $\epsilon = 1$ can be treated as the phase when the universe is just about to start its decelerating phase.}. The number of e-folds, namely, $N(H)$, could be calculated using
the formula $N(H)=-\int_{H_{end}}^H\frac{H}{\dot{H}}dH$, leading to
\begin{eqnarray}
N(x)=\frac{1}{x}-\frac{1}{x_{end}}+\frac{1}{12}  \left[ \ln\left(\frac{12-x}{12-x_{end}} \right) + \ln  \left (\frac{x_{end}}{x} \right)  \right],
\end{eqnarray}
where $x_{end}$ is the value of the parameter $x$ when inflation ends, and $x_{end}=6(1-\sqrt{2/3})\cong 1.1010$. The values of $x$ which allow to fit well with the theoretical values of the 
inflationary parameters, namely,
the spectral index, its running and the tensor/scalar ratio with their observable values,
we  obtain $64\leq N(x)\leq 117$.

To determinate the value of $\xi_0$, one has to take into account the theoretical \cite{btw} and the observational \cite{bld} value of the power spectrum
${\mathcal P}\cong \frac{H^2}{8\pi^2\epsilon}=\frac{9\xi_0^2}{32\pi^2\epsilon x^2}=\frac{9\xi_0^2}{4\pi m^2_{pl}\epsilon x^2}\cong 2\times 10^{-9}$,
where we have  explicitly introduced the Planck's mass,
which in our units is $m_{pl}=\sqrt{8\pi}$. Using the values of $x$ in the range
$[0.0085,0.0156]$, one easily finds that, $4\times 10^{-8}{m_{pl}}\leq \xi_0\leq 10^{-7}{m_{pl}}$.
Thus, summing up, the observable modes that in our model leave the Hubble radius at scales
$6\times 10^{-12} \rho_{pl}\lesssim \rho=\frac{ 3H^2m_{pl}^2}{8\pi} \lesssim   10^{-11} \rho_{pl}$,
where $\rho_{pl}=m_{pl}^4$ is the Planck's energy density leading to a power spectrum that fit well with current observable data.\newline


Let us now consider the gravitational particle production. Due to an abrupt phase transition, the production of light $\chi$-particles
become nearly conformally coupled with gravity \cite{ford}.
Their energy density will be given by
\cite{Birrell}
\begin{eqnarray}
 \rho_{\chi}=\frac{1}{(2\pi a)^3 \,a}\int_0^{\infty}k|\beta_k|^2 d^3k,
\end{eqnarray}
where the frequency $\omega_k(t)=\sqrt{k^2+M^2 a^2 (t)}$, of the produced particles in the $k$-mode
has been approximated by $k$, because we are dealing with light particles ($M\ll 1$), and the $\beta_k$-Bogoliubov coefficient is given by \cite{Birell1, Zeldovich}
\begin{eqnarray} \label{beta}
 \beta_k\cong \frac{i(\tilde{\xi}-\frac{1}{6})}{2k}\int_{-\infty}^{\infty}e^{-2ik\tau}a^2(\tau)
 R(\tau) d\tau,
\end{eqnarray}
where $R=6(\dot{H}+2H^2)$ is the scalar curvature, $\tau$  is the conformal time,
$\tilde\xi$ being the coupling constant, and we have chosen $\xi_1=0$ in order to ensure the convergence of (\ref{beta}).

Applying integrating by parts twice in Eq. (\ref{beta}), we have $\beta_k\sim {\mathcal O}(k^{-3})$, this means the energy density of the produced particles is not ultra-violet divergent 
(this is due to the fact that the coefficient of bulk viscosity is continuous during the phase transition \cite{ford}). Moreover,
 $\beta_k=(\tilde{\xi}-\frac{1}{6})f(\frac{k}{a_E\xi_0})$, where $f$ is some function, and $a_E$ is the value of the scale factor at the phase transition time.
Then, choosing for instance \footnote{Let us make a note that we are dealing with nearly conformally coupled particles.}, $\tilde{\xi}-\frac{1}{6}\sim 10^{-1}$, one obtains that 
the energy density of the produced particles is of the order $\rho_{\chi}
\sim 10^{-2}{\mathcal N}\xi_0^{4}\left(\frac{a_E}{a}\right)^4$, 
where ${\mathcal N}\equiv \frac{1}{2\pi^2}\int_0^{\infty}s^3f^2(s)ds$, is a dimensionless quantity of order $1$ .

After the phase transition, at first,
these particles will interact with each other exchanging gauge bosons and constitute a relativistic plasma that
thermalises our universe \cite{spokoiny,pv} before it was radiation
dominated. Further, the background
in our model in a deflationary stage, means that the energy density decays as $a^{-6}$,
and on the other hand, the energy density of the produced particles 
decreases as $a^{-4}$.
Eventually, the energy density of the produced particles must dominate and the
universe will become radiation dominated which matches with the hot Friedmann universe.
The universe will expand, as well as, it will cool, as a result, the
particles will be non-relativistic in nature, and hence, the universe enters into a matter dominated phase,
essential for the grow of cosmological perturbations,
and finally, only at very late time,
when  $H\sim\xi_1$, scalar field comes back to start the cosmic acceleration.

The reheating temperature, $T_R$,
is defined as the temperature of the universe when the energy density of the background and the energy density the produced particles are of the same order
 ($\rho\sim \rho_{\chi}$). Now, as $\rho_{\chi}\sim 10^{-2}{\mathcal N} \xi_0^4\left(\frac{a_E}{a}\right)^4$, and $\rho\sim 7\times 10^{-3} \xi_0^2m_{pl}^2\left(\frac{a_E}{a}\right)^6$, thus,
 one may find $\frac{a_E}{a(t_R)}\sim 10^{-1}\sqrt{\mathcal N}\frac{\xi_0}{m_{pl}}$, and hence,
one finds that $T_R\sim \rho_{\chi}^{1/4}(t_R)\sim  10^{-1}{\mathcal N}^{\frac{3}{4}}\frac{\xi_0^2}{m_{pl}}\sim 10^3{\mathcal N}^{\frac{3}{4}}\mbox{ GeV } \sim 10^3\mbox{ GeV }$,
which is below the GUT scale ($10^{16}$ GeV), meaning that the GUT symmetries are not restored
preventing a second  monopole production stage.
Moreover, this also guaranties the standard successes with
nucleosynthesis, because it  requires a
reheating temperature  below $10^{9}$ GeV \cite{37.1, 37.2}.

Lastly, we follow \cite{pv, spokoiny} to calculate temperature at the equilibrium
phase.
The interaction rate, $\Gamma$, is given by $\Gamma=n_{\chi}\sigma$, where  the cross section of scattering is
$\sigma\sim \frac{\alpha^2}{\bar{\epsilon}^2}$ where $\alpha$ is a coupling constant, and $\bar{\epsilon}\sim H_E\left(\frac{a_E}{a} \right)$ is the typical energy of a produced particle, and the energy density of the produced particles is given by \cite{Birrell}
\begin{eqnarray}
 \hspace{-0.35cm} n_{\chi}=\frac{1}{(2\pi a)^3}\int_0^{\infty}|\beta_k|^2 d^3k\sim
  10^{-2}{\mathcal M}\xi_0^{3}\left(\frac{a_E}{a}\right)^3,
\end{eqnarray}
where $ {\mathcal M}\equiv \frac{1}{16\pi}\int_{-\infty}^{\infty} a^4(\tau)R^2(\tau) d\tau$, is also a dimensionless quantity of order $1$. 

Now, since the thermal equilibrium is achieved when $\Gamma\sim H(t_{eq})=H_E\left(\frac{a_E}{a_{eq}}\right)^3$ (Recall that, for our model, this process is produced in the still fluid era),
and when the equilibrium is reached one has
$\frac{a_E}{a_{eq}}\sim \alpha {\mathcal M}^{\frac{1}{2}}$, hence, the equilibrium temperature is found to be
$T_{eq}\sim {\mathcal N}^{\frac{1}{4}}{\mathcal M}^{\frac{1}{2}}\alpha \xi_0$. Therefore, we have
$T_{eq}\sim (10^{11}-10^{12}){\mathcal N}^{\frac{1}{4}}{\mathcal M}^{\frac{1}{2}}\alpha \mbox{ GeV}.$
Choosing as usual $\alpha\sim (10^{-2}-10^{-1})$ \cite{spokoiny, pv}, one has the following equilibrium temperature,
$T_{eq}\sim (10^9- 10^{11}){\mathcal N}^{\frac{1}{4}} {\mathcal M}^{\frac{1}{2}}\mbox{ GeV } \sim (10^9- 10^{11})\mbox{ GeV }.$

\section{Thermodynamics of Bulk viscous cosmology}
\label{thermo}

Thermodynamics plays a crucial role in investigating any 
cosmological model. The viabilities of any cosmological model
depends on its thermdoynamical properties. To investigate the 
themrodynamical laws associated to any cosmological model, one
assumes the universe to be a thermodyamical system bounded by some
cosmological horizon and the total matter of the universe is
enclosed within a comving volume of a radius not exceeding 
the cosmological horizon mentioned above. The idea of such 
cosmological horizon was originated from the black hole 
thermdoynamics, and interestingly, the thermodynamical properties that hold for a black hole are also valid for a  cosmological horizon \cite{gibbons, jacobson, paddy}. 
Moreover, it is quite riveting that the first law of thermodynamics which holds for the black hole horizon can be derived from the first Friedmann equation for the FLRW geometry in which the universe is bounded by an apparent horizon. In what follows, 
the choice of an apparent horizon as the cosmological horizon provides a good motivation to investigate the thermdoynamical laws for any 
cosmological model under consideration. The radius of the apparent horizon for the FLRW universe can be calculated as $r_h= \left(H^2+ k/a^2  \right)^{-1/2}$ \cite{Bak}.  Thus, for $k= 0$ (the flat FLRW universe) the radius becomes $r_h = 1/H$, which is the so-called Hubble horizon. Now, the natural tendency of systems, here the cosmological model, is to evolve toward an thermodynamic equlibrium phase is governed by its entropy,
$S$, where the entropy should satisfy the following two properties, namely that the entropy of the system should be non-decreasing with the evolution of the universe, that means, $\dot{S} \geq 0$ (the overdot represents the cosmic time differentiation), and it is known as the second law of thermodynamics. Secondly, the entropy should be convex, that is, $\ddot{S} <0$, which is the condition for equilibrium \cite{callen}. 
The entropy is contributed from the entropy of
the apparent horizon and the entropy of the fluid 
bounded by the horizon. So, $S= S_h+ S_f$, where $S_h$, $S_f$
denote the entropies of the apparent horizon and the fluid respectively.
The entropy of the apparent horizon is found to be $S_h = k_{B} \mathcal{A}/4\, l_{pl}^2$,
where $k_{B}$ is the Boltzmann's constant, $l_{pl}= \sqrt{1/8 \pi}$ is the Planck's length in our units; $\mathcal{A}= 4 \pi r_h^2$, is the horizon area. We assume that
temperature of the apparent horizon 
is Hawking temperature which is $T_A= 1/2\pi r_h= H/2\pi$ \cite{temperature1, temperature2, temperature3, temperature4, temperature5, temperature6}.
Now, for the present bulk viscous model, since we found that the 
model has a nonsigular structure together with two successive 
accelerating phase, thus, it will be worth to examine  
whether the model attains an equlibrium phase in its long run.
In order to do so, let  us differentiate $S_h$ with respect to the 
cosmic time and using the Friedmann and Raychaudhuri 
equations, one arrives at

\begin{align}\label{th1}
\dot{S}_h & = -\, \left(\frac{2 \pi k_B}{l_{pl}^2\, H^3}\right)\, \dot{H}  = \frac{3 \pi k_B}{l_{pl}^2 H}\, \gamma\, \left(1-\frac{\xi (H)}{\gamma H} \right),
\end{align} 
Now, we recall the Gibb's equation for the fluid which relates the thermodynamic quantities associated with the fluid in the following way
\begin{align}\label{gibbs}
T dS_{f} &= d (\rho\, V) + p dV,
\end{align}
where  $V= 4\pi r_h^3/ 3$ is the volume of the region surrounded by the radius $r_h$,
and $T$ is the fluid temperature which is equal to $T_A$. We note that the temperature of the fluid should be equal to that
of the horizon temperature so that no effective flow of the fluid is found toward
the horizon. Now, introducing the cosmic time in the above equation (\ref{gibbs})

\begin{align}\label{fluid-entropy}
\dot{S}_{f} &= -8 \pi^2\, (3\gamma - 2)\, \frac{\dot{H}}{H^2} =\frac{12\,\gamma\,\pi^2}{H}\,\left(1-\frac{\xi(H)}{\gamma H}\right)(3\gamma-2).
\end{align}
Since $ 1 \leq \gamma \leq 2$, we must have that $3\gamma- 2 > 0$. Since in the quintessence era, $\dot{H}< 0$, thus it is easily seen that $\dot{S}_h> 0$, and $\dot{S}_f > 0$. Hence, $(  \dot{S}_h+ \dot{S}_f  ) > 0$, which implies that the entropy is always increasing with the increase of the cosmic time.  Now let us consider the second order derivatives of the entropies in order to see the equilibrium condition. Now, differentiating (\ref{th1}) with respect to the cosmic time, we find 

\begin{equation}\label{bv1}
\ddot{S}_h= \frac{3\pi k_B}{l_{pl}^2}\, \frac{\dot{H}}{H^3}\,\Bigl(\gamma H- H \xi^{\prime} (H)- 2 \gamma H + 2 \xi (H)\Bigr)= \frac{3\pi k_B}{l_{pl}^2}\, \frac{\dot{H}}{H^3}\, \Bigl( (m-\gamma) H - 2\xi_0 + \frac{3n}{H}  \Bigr),
\end{equation}
and similarly, differentiating (\ref{fluid-entropy}) using the cosmic time, one gets 

\begin{equation}\label{bv2}
\ddot{S}_f = 12 \pi^2 \left(3\gamma-2  \right ) \frac{\dot{H}}{H^3}\,\Bigl(\gamma H- H \xi^{\prime} (H)- 2 \gamma H + 2 \xi (H)\Bigr)= 12 \pi^2 \left(3\gamma-2  \right )\, \frac{\dot{H}}{H^3}\, \Bigl( (m-\gamma) H - 2\xi_0 + \frac{3n}{H}  \Bigr).
\end{equation}
Now, one can check the condition for equilibrium for this model from the condition $\ddot{S} = \ddot{S}_h+ \ddot{S}_f$. Thus,  using the previous two equations (\ref{bv1}) and (\ref{bv2}), one finds that

\begin{eqnarray}\label{bv1+bv2}
 \ddot{S}_h+ \ddot{S}_f = \left( \frac{3\pi k_B}{l_{pl}^2} + 12 \pi^2 \left(3\gamma-2  \right ) \right)  \frac{\dot{H}}{H^3}\, \Bigl( (m-\gamma) H - 2\xi_0 + \frac{3n}{H}  \Bigr)
\end{eqnarray}
One may notice that since $ 1 \leq \gamma \leq 2$, thus the quantity $\left( \frac{3\pi k_B}{l_{pl}^2} + 12 \pi^2 \left(3\gamma-2  \right ) \right)$ is always positive. Hence,  the equilibrium condition is determined by the quantity  $\left[ (m-\gamma) H - 2\xi_0 + \frac{3n}{H}  \right]$. Now, let us calculate the quantities in equations (\ref{bv1}) and (\ref{bv2}) near the attractor ($H_{-}$) and the repeller ($H_{+}$) critical points. It is easy to see that

\begin{equation}\label{xxx}
\Bigl( (m-\gamma) H - 2\xi_0 + \frac{3n}{H}  \Bigr)_{H= H_{-}}= \xi_0\, \sqrt{1+ \frac{4 (\gamma- m)\, n}{\xi_0^2}} > 0,\,\,\,\,\text{as}\,\, (\xi_0, m, \gamma, n) \in W
\end{equation}
and 
\begin{equation}
\Bigl( (m-\gamma) H - 2\xi_0 + \frac{3n}{H}  \Bigr)_{H= H_{+}}= - \xi_0\, \sqrt{1+ \frac{4 (\gamma- m)\, n}{\xi_0^2}} < 0,\,\,\,\,\text{as}\,\, (\xi_0, m, \gamma, n) \in W
\end{equation}

Thus, considering (\ref{bv1+bv2}) for the quantities given in equations (\ref{bv1}) and (\ref{bv2}) for $\dot{H}< 0$ (equivalently, the non-phantom fluids, see the Raychaudhuri equation (\ref{friedmann2})) one can derive the equilibrium conditions for the current bulk viscous model at the fixed points $H_{-}$ (attractor) and $H_{+}$ (repeller) and these lead to the following results:

\begin{equation}
\Bigl(  \ddot{S}_h+ \ddot{S}_f \Bigr)_{H \rightarrow H_{-}} < 0\,\,\,\,(\mbox{Equilibrium condition}),
\end{equation}
and

\begin{equation}
\Bigl(  \ddot{S}_h+ \ddot{S}_f \Bigr)_{H \rightarrow H_{+}} > 0\,\,\,\,(\mbox{Non-equilibrium condition}).
\end{equation}
That means, the model tends toward a thermodynamic equilibrium phase as $H \rightarrow H_{-}$  while for $H \rightarrow H_{+}$, the model shows the non-equilibrium tendency.

Now, we introduce the 
quantum corrections to Bekenstein-Hawking entropy 
law which enables us to  generalize the 
black hole horizons as \cite{MP13}

\begin{align}\label{bv01}
S_h & = k_B \left[ \frac{\mathcal{A}}{4 l_{pl}^2} - \frac{1}{2} \ln \left(\frac{\mathcal{A}}{l_{pl}^2} \right)\right],
\end{align}
including some higher order corrections \cite{new1, new2}. We assume that this definition holds good to the cosmic apparent horizon \cite{MP13}, and one may check how the change in entropy occurs, that means now whether the equlibrium condition is reached or not. Thus, using one time cosmic time differentiation in eqn. (\ref{bv01}) we have 

\begin{equation}
\dot{S}_h = \frac{k_B}{l_{pl}^2} \frac{\dot{H}}{H}\, \left( -\frac{2 \pi}{H^2}+ l_{pl}^2\right).
\end{equation}
Now, since $\dot{H}< 0$, then $\dot{S}_h$ is positive for $H < \frac{\sqrt{2 \pi}}{l_{pl}}$, that means when the universe leaves the very early phase (i.e. Planck stage) and enters into the clasical stage, then $\dot{S}_h > 0$, i.e. entropy is increasing implies the validity of the first law of thermodynamics. Now, the second derivative of entropy gives 

\begin{eqnarray}
\ddot{S}_h = -\frac{3\gamma k_B}{2 l_{pl}^2}\, \dot{H} \left[\left(1-\frac{\xi^\prime (H)}{\gamma}\right) \left(-\frac{2\pi}{H^2}+ l_{pl}^2\right) + \left(H- \frac{\xi(H)}{\gamma}\right) \frac{4\pi}{H^3}\right]\nonumber\\
= -\frac{3\gamma k_B}{2 l_{pl}^2}\, \dot{H} \left[l_{pl}^2+ \frac{2\pi}{H^2} \left(1-\frac{2 \xi(H)}{\gamma H}\right)+ \frac{\xi^\prime (H)}{\gamma} \left(\frac{2\pi}{H^2}- l_{pl}^2\right)\right]
\end{eqnarray}
Now, near the attractor point, the above expression can be approximated as 

\begin{eqnarray}
\ddot{S}_h (H \rightarrow H_{-})
\simeq \frac{3 \pi \gamma k_B \dot{H}}{ l_{pl}^2\, H_{-}^4}\,  \Bigl[ (m-\gamma) H_{-}^2 - 2 \xi_0 H_{-} + 3n\Bigr] 
\end{eqnarray}
which is negative since $\dot{H}< 0$ and the quantity inside the third brace is positive that follows from (\ref{xxx}). That means, near the attarctor point, entropy is convex. This implies the present bulk viscous model is in agreement with the thermodynamics of the universe, and even the laws hold when the Bekenstein-Hawking correction is also accounted.
So, the choice of $\xi (H) = \xi_0 + m H + \frac{n}{H}$, is consistent with the thermodynmical laws and the universe asymptotically attains the thermodynmaical equilibrium.

\section{Summary}
\label{summary}

In the context of general relativity, in this work, we have discussed the cosmological solutions for a spatially  flat FLRW universe driven by some bulk viscous pressure where the background matter has a perfect fluid description with non-negative pressure. Such a cosmological scenario has an equivalent description in terms of an inhomogeneous equation of state as discussed in some earlier works \cite{Nojiri:2005sr, Capozziello:2005pa, Nojiri:2006zh, bcno}. The cofficient of bulk viscosity, $\xi(t)$, could in general be a function of the scale factor, Hubble parameter of the FLRW universe and also the derivatives of the Hubble parameter, in principle. Thus, there is no such strict restriction on the choice of the bulk viscous models. In fact, it has been found that for different phenomenological choices for $\xi (t)$, the early inflationary \cite{Barrow:1986yf, Barrow:1988yc, bo, bt, Brevik:2016kwq} and late cosmic acceleration \cite{beno, begt, bcno} can be explained. Thus, the search for a new  
bulk viscous model unifying both the accelerating regimes is indeed an interesting idea in this direction. In order to do so we propose a simple bulk  viscous model that only depends on teh FLRW Hubble rate $H$  with the expression $\xi (H) = -\xi_0 + mH + n/H$ (where $(\xi, m, n) \in \mathbb{R}^3$). This model is interesting in the sense that it recovers some simple models with $\xi (H) = $ constant, $\xi (H)\propto H$ and $\xi (H) \propto 1/H$, as special cases. We solved the Einstein's field equations for this general bulk viscous model and performed a dynamical systems analysis. From our analysis,  we found that the model predicts two distinct accelerated phases, one at very early time and other is the current observed accelerating phase. However, the early and the late accelerating phases are represented by two critical points respectively with repeller and attractor behavior, that means the universe leaves the early accelerating phase and reaches the current one. Additionally, the model is singularity 
free. Further, it is interesting to note that the current bulk viscous model unifying the accelerated expansions can be mimicked by a single scalar field with a
potential which is perfectly analytic and it is a combination of a one dimensional Higgs potential (that explains the early inflationary expansion) plus an exponential potential which responsible for the current cosmic acceleration.  
The potential leads to a power spectrum of the cosmological
perturbations which shows that if the running is disregradred then the tensor-to-scalar ratio, $r$, for the present model is slightly high ($0.1344\leq r \leq 0.3072$) in respect to the estimations from Planck 2013 ($r <0.11$ at 95.5\% CL) \cite{Ade} and  Planck 2015 ($r< 0.12$ at 95.5\% CL) \cite{Ade:2015lrj}, but when the running is allowed, the tensor-to-scalar ratio for the model is constrained to be $r \leq 0.25$ at 95.5\% CL \cite{Ade}. 
After the early accelerating phase, an abrupt phase transition occurs due to
gravitational particle production and the universe is thermalized as a result of this. Consequently, the universe becomes radiation dominated, and due to its
expansion it automatically cools down and enters into the non-relativistic matter era. Finally, we observe the current acceleration of the universe. We show that the current bulk viscous model is consistent with the thermodynamics, that means the entropy of the universe is always increasing and it tends toward a thermodynamic equilibrium in the late de Sitter point which is expected. In fact, the thermodynamics of this bulk viscous scenario is also consistent when the Bekenstein-Hawking correction is also introduced. 

Thus, the present bulk viscous model provides with a thermodynamically consistent cosmological scenario which is singularity free and produces two successive accelerating phases, namely, the early inflationary (dynamically unstable) and late (dynamically stable) accelerating phases where the model allows an intermediate hot radiation and cold matter dominated eras due to an abrupt phase transition with gravitational particle production.


\subsection*{Acknowledgments}

The authors thank the referee for some important comments to improve the work.
We would like to thank Professor S.D. Odintsov  for his valuable comments.
The investigation of J. Haro has been supported in part by MINECO (Spain), project MTM2014-52402-C3-1-P. SP acknowledges the support from SERB-NPDF grant (File No. PDF/2015/000640). Partial support from the NBHM Post-Doctoral grant (File No. 2/40(60)/2015/R{\&}D-II/15420), Department of Atomic Energy, Govt. of India, is also acknowledged by SP.

\end{document}